\documentstyle[12pt,epsf]{article}
\newcommand{\la}[1]{\label{#1}}

\newcommand{\be}{\begin{equation}}
\newcommand{\ee}{\end{equation}}
\newcommand{\bi}{\begin{itemize}}
\newcommand{\ei}{\end{itemize}}
\newcommand{\ba}{\begin{eqnarray}}
\newcommand{\ea}{\end{eqnarray}}
\newcommand{\rmi}[1]{{\mbox{\scriptsize #1}}}

\newcommand{\nr}[1]{(\ref{#1})}

\newcommand{\bfs}{\mbox{\bf s}}

\newcommand{\bfb}{\mbox{\bf b}}

\newcommand{\roots}{\sqrt{s}}

\newcommand{\gev}{{\rm GeV}}
\newcommand{\psat}{p_{\rmi{sat}}}
 
\newcommand{\qsat}{Q_{\rmi{sat}}}
\newcommand{\lQCD}{\Lambda_{\rm QCD}}

\newcommand{\fr}[2]{{\frac{#1}{#2}}}

\def\lsim{\,\raise0.3ex\hbox{$<$\kern-0.75em\raise-1.1ex\hbox{$\sim$}}\,}
\def\gsim{\,\raise0.3ex\hbox{$>$\kern-0.75em\raise-1.1ex\hbox{$\sim$}}\,}

\begin{document}
  
\begin{titlepage}
\begin{flushright}
29 June 2001\\
JYFL-7/01\\
HIP-2001-26/TH\\
hep-ph/0106330\\
\end{flushright}
\begin{centering}
\vfill

{\bf 
HEAVY ION COLLISION MULTIPLICITIES AND GLUON DISTRIBUTION FUNCTIONS
}

\vspace{0.5cm}
 K.J. Eskola,$^{\rm a,c,}$\footnote{kari.eskola@phys.jyu.fi}
 K. Kajantie$^{\rm b,}$\footnote{keijo.kajantie@helsinki.fi}  and
K. Tuominen$^{\rm a,}$\footnote{kimmo.tuominen@phys.jyu.fi}

\vspace{1cm}
{\em $^{\rm a}$ Department of Physics, University of Jyv\"askyl\"a,\\
P.O.Box 35, FIN-40351 Jyv\"askyl\"a, Finland\\}
\vspace{0.3cm}
{\em $^{\rm b}$ Department of Physics, University of Helsinki\\
P.O.Box 64, FIN-00014 University of Helsinki, Finland\\}
\vspace{0.3cm}
{\em $^{\rm c}$ Helsinki Institute of Physics,\\
P.O.Box 64, FIN-00014 University of Helsinki, Finland\\}

\vspace{1cm}
{\bf Abstract}

\end{centering}

\vspace{0.3cm}\noindent
Atomic number ($A$) and energy ($\roots$) scaling exponents of
multiplicity and transverse energy in heavy ion collisions are
analytically derived in the perturbative QCD + saturation model. The
exponents depend on the small-$x$ behaviour of
gluon distribution functions at an $x$-dependent scale. 
The relation between initial
state and final state saturation is also discussed.

\vfill 

\end{titlepage}

\section{Introduction}

New RHIC data on the multiplicity in Au+Au collisions
\cite{phobos1} and its centrality dependence \cite{phenix,phobos2} has given 
us new insight into the dynamics of ultrarelativistic heavy ion
collisions. The data from central and nearly central collisions 
can be understood \cite{wg,kharzeevnardi}
in terms of a conventional soft + hard two-component
picture \cite{ekl}, but also in a dynamically more unified 
perturbative QCD + saturation model \cite{ekrt,ekt}. 

The results of \cite{ekrt} are formulated in the form of scaling
rules: quantity $\sim CA^a(\roots)^b$, where the constants $C,a,b$ are
determined numerically for central A+A collisions using independently
determined parton distribution functions \cite{grv} with shadowing
\cite{EKS98}.  For example, for the dominant saturation scale $\psat$
and for the multiplicity $N$ per unit rapidity one finds: 
\ba
\psat&=&0.208\,A^{0.128}\,\roots^{\,0.191}, \la{psatfit}\\
N(\psat)&=&1.383\,A^{0.922}\roots^{\,0.383}, \la{npsatfit} 
\ea 
where $\psat$ and $\roots$ are in units of GeV.  Thus the multiplicity
$N\sim A^{0.922}$ instead of the exponent $N\sim A^{4/3}$ appropriate
for hard collisions or $N\sim A$ appropriate for the saturation model
if $xg(x)\sim$ const and $\alpha_s\sim$ const. Even more striking is the
rather fast powerlike dependence $N\sim\roots^{0.383}$, much faster
than the $\sim\log(s)$ or $\sim\log^2(s)$ behaviour observed for pp
collisions.

The results in \nr{psatfit}-\nr{npsatfit} lead to rather definite
and easily testable 
predictions for the overall magnitude and the $A$ and
$\roots$ dependencies of heavy ion experimental results. Agreement
with the first RHIC results for the charged multiplicity and 
$\roots$ dependence is good \cite{phobos1}.
It would thus be of some value to
derive the numerically computed scaling parameters analytically
and to understand the underlying physics. It is
the purpose of this note to carry out this derivation.

The dominant part of the analytic estimate is a derivation of 
an accurate approximation for hard production of minijets in pp
collisions. This, of course, is a most standard problem, but with
one important difference: as we are interested in the physics of
heavy ion collisions in the RHIC-LHC energy range, $\roots\gsim100$ GeV, 
and for $A\sim200$, we know the magnitudes of the dominant scales of
the problem, $Q\sim \psat$ and $x\sim \psat/\roots$. It is thus
sufficient to find out an approximation for the gluon distribution
functions in this range. In fact, we shall find that the very
simple estimate
$xg(x,Q^2)\sim (Q/x)^{\delta}$ with $\delta\approx 0.5$ 
is quite accurate in the vicinity of the 
dominant scale $Q$, which depends on $x$. 
After obtaining an analytic approximation
for the perturbative minijet cross section
$d\sigma/dy(y=0, p_T\ge p_0)$ (Section 2) it is 
straightforward to find the scaling exponents (Section 3).

The idea of
saturation originates from \cite{glr,mq,bm} and is usually discussed
in connection with small-$x$ behaviour of gluon distribution
functions (``initial state saturation"). 
Here we have in mind rather a picture with
the saturation of produced gluons (``final state saturation"). Using 
the analytic approximation we can show the close phenomenological
relation of initial and final state saturation (Section 4).
This close relation also follows from the fact 
that $N(\psat)$ is proportional to the initial state gluon
distribution probed at the final state saturation scale. The
analytic approximation can also be extended to models of
local saturation (Section 5).

\section{Analytic estimates of the minijet cross section}

Consider first inclusive gluon production from the subprocess
gg$\to$gg in pp collisions: 
\be 
{d\sigma\over dy d^2p_T}=K\int dy_2\,x_1g(x_1,p_T^2)\,x_2g(x_2,p_T^2)
{9\alpha_s^2\over2p_T^4}\left(1-{x_T^2\over 4x_1x_2}\right)^3.
\la{incl} 
\ee 
Here $K$ describes the effect of higher order
corrections \cite{et} and the fractional momenta are
$x_1=\fr12x_T(e^y+e^{y_2}),\,x_2=\fr12x_T(e^{-y}+e^{-y_2})$ with
$x_T=2p_T/\roots$. The integral is over $-\log(2/x_T-e^{-y})\le y_2\le
\log(2/x_T-e^y)$.  The last factor in \nr{incl} is equal to
$(1+2\cosh Y)^3/(2+2\cosh Y)^3,\,\, Y=y-y_2$.

For the analytic estimates, we need to approximate the gluon distribution 
$xg(x,Q^2)$ in a region which dominates the $p_T$- and $y_2$-integrations. 
From \nr{npsatfit}, we see that $\psat\sim 1$ GeV at
$\roots=200$ GeV, and $\psat\sim 2$ GeV at $\roots=5500$ GeV for
$A\sim200$.  In addition, as shown in Fig. 1 of \cite{ekrt}, we know
that about 90\% of all the minijets produced above the saturation
scale have transverse momenta $\psat \le p_T \,\lsim\, 2 \psat$.  The
scale in \nr{incl} is chosen as $Q=p_T$, so the relevant region in
$Q$ thus is $1\dots2$ GeV for $\roots=200$ GeV and $2\dots4$ GeV for
$\roots=5500$ GeV. In what follows, we shall fix $y=0$ in
\nr{incl}. This makes the integral even and it suffices to consider
only the integration region $0\le y_2 \le \log(2/x_T-1)$. The
fractional momenta are now limited to $x_T/2\le x_1\le 1$ and
$x_T/2\approx(2/x_T-1)^{-1}\le x_2\le x_T/2$. From the numerical
computation of \nr{incl}, we have checked
that the dominant ($\ge 70$\%) 
contribution comes from $y_2\lsim 4$ for $\roots=5500$
GeV ($p_T\ge 2$ GeV), and from $y_2\lsim3$ for $\roots=200$ GeV ($p_T\ge 1$
GeV). In the analytic estimates we thus need to describe $xg(x,Q^2)$ at
$Q/\roots\lsim x \lsim 50 Q/\roots$ in the vicinity of $Q=2$ GeV for
$\roots=5500$ GeV and at $Q/\roots\lsim x\lsim 20 Q/\roots$ in the
vicinity of $Q=1$ GeV for $\roots=200$ GeV.

\begin{figure}[hbt]
\vspace{-3cm}
\hspace{0cm}
\centerline{\hspace{1cm}\epsfysize=9.5cm\epsffile{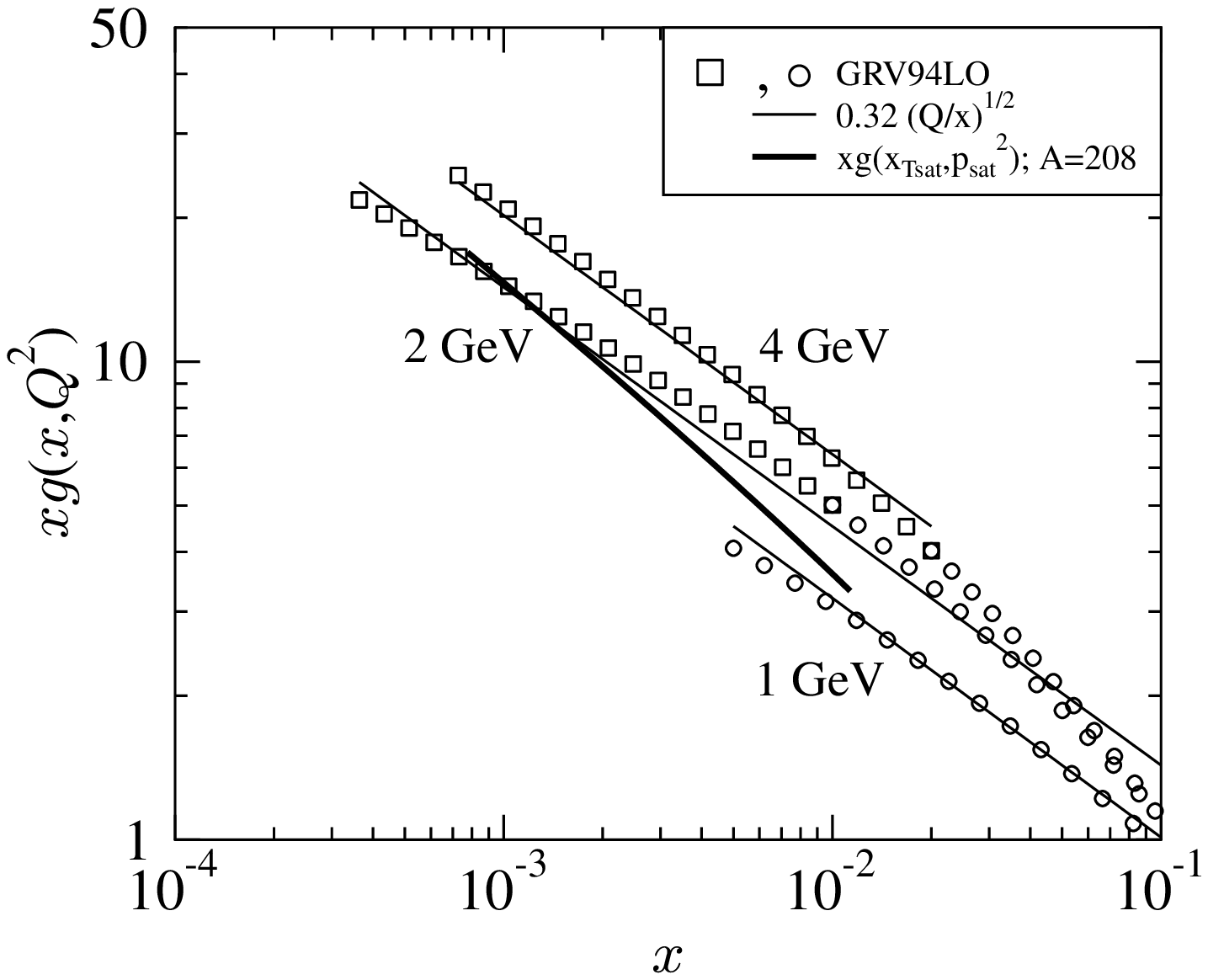} 
\hspace{-1cm}\epsfysize=9.5cm\epsffile{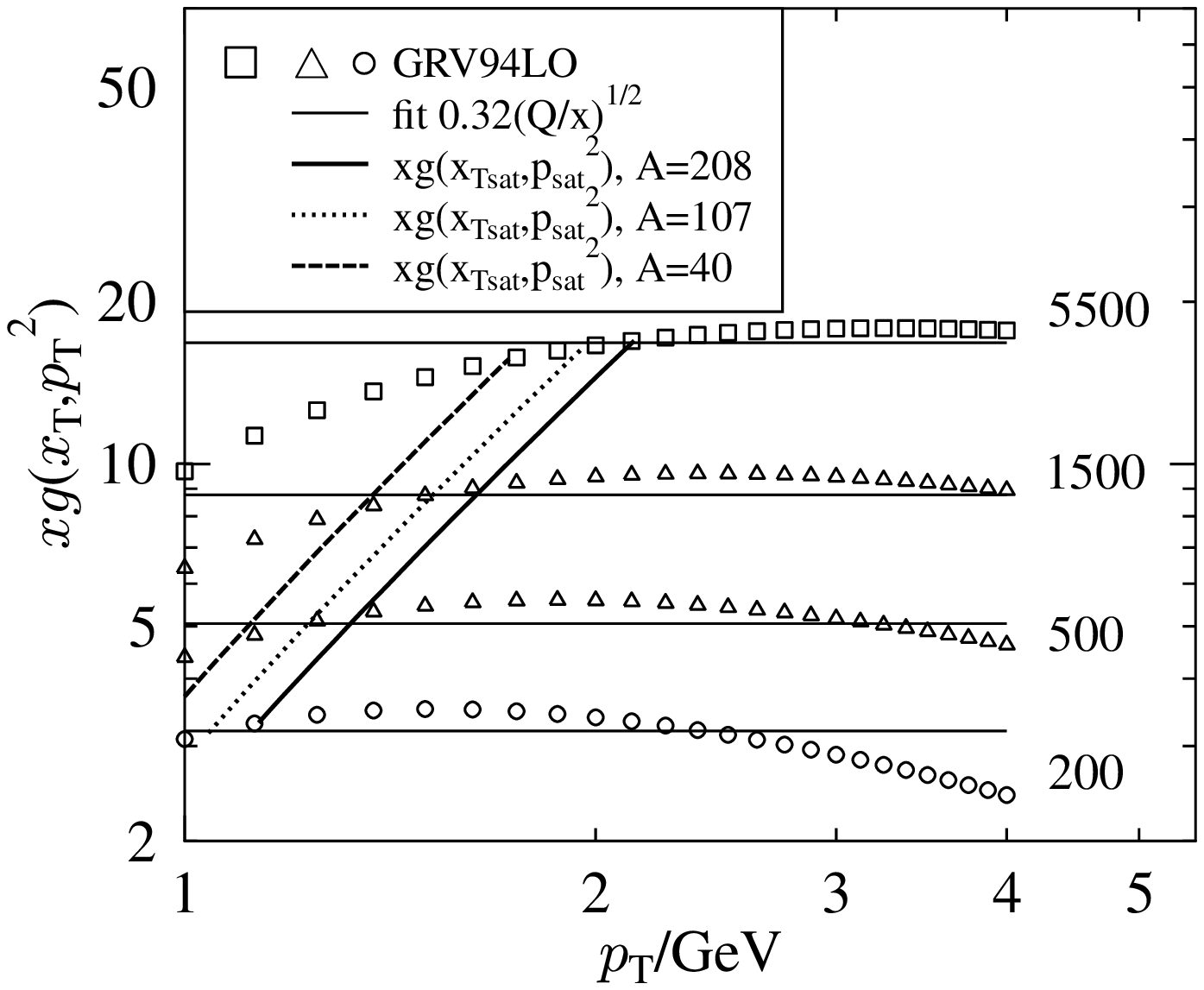} }
\vspace{-1cm}
\caption[a]{\small (a) The gluon distribution $xg(x,Q^2)$ as a
function of $x$ at fixed scales $Q$, plotted in the
region which dominates minijet production at saturation at
$\roots=200...5500$ GeV, i.e. at $x\gsim Q/\roots$ 
with $Q=1$, 2 and 4 GeV (see the text for details).
The squares and circles show the GRV94-LO distributions \cite{grv},
the solid lines are the fit $xg(x,Q^2)=0.32(Q/\gev/x)^{0.5}$. 
The solid thick
line shows the gluon density probed at saturation, $xg(x_{T{\rm
sat}},\psat^2)$ with $A=208$ at $\roots=200\dots 5500$ GeV.  (b) The
gluon distribution $xg(x_T,p_T^2)$ as a function of the scale $p_T$
for $\roots=200$, 500, 1500 and 5500 GeV. The symbols are the GRV
distributions and the thin solid lines are the fit. The thick curves
(solid, dotted, dashed) are the gluon densities probed at saturation,
$xg(x_{T{\rm sat}},\psat^2)$, at different $\roots$ for $A=208$, 107
and 40, correspondingly.  The region relevant for our problem is to
the right of the thick curves.  } \la{viuhka}
\end{figure}

The gluon distribution $xg(x,Q^2)$ as obtained from the set GRV94-LO
\cite{grv} in the $x,Q$ region discussed above is
shown in Fig.1a. The symbols are for $\roots =200$ GeV (circles) and
for $\roots =5500$ GeV (squares). For our purposes, and as the
kinematic region is now limited, a simple power law $x^{-0.5}$, shown
by the solid lines in Fig.1a, reproduces $xg$ adequately. To simulate
the effect of pQCD scale evolution, we note from Fig.1b that in the
dominant region (to the right of the thick tilted lines), $xg(x\sim
p_T/\roots,p_T^2)$ is approximately constant. This suggests a simple
fit 
\be
xg(x,Q^2)=C_0(\frac{Q/\gev}{x})^{\delta}, 
\la{simplefit}
\ee
with $\delta=0.5$ and 
$C_0=0.32$,  as shown by the solid lines in Fig.1.
Below we shall see that $N(\psat)\sim xg(x_{T{\rm sat}},\psat^2)$,
with $x_{T{\rm sat}}=2\psat/\roots = 0.416 A^{0.128}(\roots/\gev)^{-0.809}$,
where the relation \nr{psatfit} is used. The gluon density probed at
final saturation, $xg(x_{T{\rm sat}},\psat^2)$ is shown by the thick 
line in Fig. 1a for $A=208$ from $\roots=200$ GeV to 5500 GeV.
Note that on this curve each point in $x$ now corresponds to 
one $\roots$ and one $\psat$. 

If a similar procedure is carried out for the CTEQ5 set of parton
distribution functions \cite{cteq5}, effectively the same result is
obtained, the value of $C_0$ is decreased by less than 10\% and
$\delta\approx0.47$ is somewhat smaller.

With the fit $xg(x,p_T^2)=C_0 (p_T/{\rm GeV}/x)^{\delta}$, 
($\delta=0.5$, $C_0$=0.32),
Eq. \nr{incl} can now be expressed as 
\be 
\frac{d\sigma}{dy d^2p_T} =
2\cdot K \frac{9\alpha_{\rm s}^2(p_T^2)}{2 p_T^4}
[C_0(\frac{\roots}{\gev})^{\delta}]^2 \sum_{n=0}^3 (-1)^n\Big({3\atop n}\Big) (\frac{1}{4})^{n+\delta}
\int_0^{{{\scriptstyle \log(\roots/p_T-1)}\atop~} \atop} 
\hspace{-2.0cm}dy_2\,\, [\cosh
\frac{y_2}{2}]^{-2(n+\delta)}. 
\ee

As our focus is at a region where $x_T\ll1$, we write
\begin{eqnarray}
\int_0^{{{\scriptstyle \log(\roots/p_T-1)}\atop~} \atop} 
\hspace{-2.0cm}dy_2\,\, [\cosh
\frac{y_2}{2}]^{-2\Delta} 
&=& \int_0^{\infty} \,dy_2 \,[\cosh \frac{y_2}{2}]^{-2\Delta}
- \int_{\atop{\atop {\scriptstyle \log(\roots/p_T-1)}}}^{\infty}
\hspace{-1.5cm}dy_2 \,[\cosh \frac{y_2}{2}]^{-2\Delta} \\ &=&
B(\Delta,\frac{1}{2}) - 2^{\Delta}\int_0^{x_T} dz
z^{\Delta-1}(1-\frac{z}{2})^{\Delta-1}\\ &=&
B(\Delta,\frac{1}{2}) +
(\frac{x_T}{2})^{\Delta}\sum_{k=0}^{\infty}\bigg({\Delta-1\atop
k}\bigg)\frac{1}{k+\Delta}(-\frac{x_T}{2})^{k},
\end{eqnarray}
where $B(\Delta,\frac{1}{2})$ is the beta function, $\Delta=n+\delta$, and 
in the second integral a change of integration variable from
$y_2$ to $z=x_T/x_1$ has been made.
In the limit of small $x_T$, the leading term for each $n$ is given by 
the first term with the beta function.  The $p_T$ distribution thus 
becomes
\be
\frac{d\sigma}{dy d^2p_T} =
2\cdot K \frac{9\alpha_{\rm s}^2(p_T^2)}{2 p_T^4}
[C_0(\frac{\roots/\gev}{2})^{\delta}]^2 B(\delta,\frac{1}{2}) a(\delta)
+ {\cal O}(x_T^{\delta})
\la{pTdist}
\ee
where 
\be
a(\delta)=1
-\frac{3}{4}\frac{\delta}{\frac{1}{2}+\delta}
+\frac{3}{4^2}\frac{\delta(\delta+1)}{(\frac{1}{2}+\delta)(\frac{3}{2}+\delta)}
-\frac{1}{4^3}\frac{\delta(\delta+1)(\delta+2)}{(\frac{1}{2}+\delta)(\frac{3}{2}+\delta)(\frac{5}{2}+\delta)}\stackrel{\scriptstyle \delta=0.5}{\approx} 0.6904
\ee
and 
\be
B(\delta,\frac{1}{2})=\frac{\sqrt\pi\Gamma(\delta)}{\Gamma(\frac{1}{2}+\delta)}
\stackrel{\scriptstyle\delta=0.5} {=}\pi
\ee
Notice that now, in \nr{pTdist}, 
$[C_0(\frac{\roots/\gev}{2})^{\delta}]^2=[xg(x_T,p_T^2)]^2$ appears.

Integration over $p_T$  then gives the minijet cross section
\ba 
2\sigma_\rmi{pQCD}(p_0) \approx {d\sigma\over dy}\bigg|_{y=0\atop p_0}
&\approx&\int_{p_0}^\infty2\pi p_Tdp_T {d\sigma\over dy d^2p_T}\\
&\approx& 
2\cdot K \frac{9\pi}{2} B(\delta,\frac{1}{2}) a(\delta) [xg(x_{T0},p_0^2)]^2 
{\cal I}(p_0^2),\la{minijetsigma}
\ea
where 
\be
{\cal I}(p_0^2)  
\equiv \int_{p_0^2}^{\infty}dp_T^2\frac{\alpha_s^2(p_T^2)}{(p_T^2)^2}
= \frac{\alpha_s^2(p_0^2)}{p_0^2}\log\frac{p_0^2}{\lQCD^2}
\bigg[ 
1-\frac{p_0^2}{\lQCD^2}\log\frac{p_0^2}{\lQCD^2}E_1(\log\frac{p_0^2}{\lQCD^2})
\bigg]
\la{Ip0}
\ee
where $E_1(z)$ is the exponential integral and the running coupling 
$\alpha_s(Q^2)$ is that to one loop.
Denoting $z=\log(p_0^2/\lQCD^2)$ and using the approximation
\cite{abramowitz} \be z{\rm e}^zE_1(z) =
\frac{z^2+a_1z+a_2}{z^2+b_1z+b_2} + \epsilon(z) \ee where
$|\epsilon(z)|< 5\cdot 10^{-5}$, $a_1 = 2.334733$, $a_2 = 0.250621$,
$b_1 = 3.330657$ and $b_2 = 1.681534$, we arrive at the following
expression for the hard cross section at central rapidity: \be
\frac{d\sigma}{dy}\bigg|_{y=0\atop p_0} \approx
2K\frac{9\pi}{2}B(\delta,\frac{1}{2}) a(\delta)\cdot
f(\log\frac{p_0^2}{\lQCD^2})\frac{\alpha_s^2(p_0^2)}{p_0^2}[xg(x_{T0},p_0^2)]^2,
\la{analytic} \ee where $f(z) =
({0.995924z+1.430913})/({z+3.330657+1.681534/z})$,
$x_{T0}=2p_0/\roots$ and
$xg(x_{T0},p_0^2)=C_0(\frac{\roots/\gev}{2})^{\delta}$.  Setting
$f(z)=1$ would correspond to the approximation 
$\alpha_s(p_T^2)=\alpha(p_0^2)$ in the integral in Eq. \nr{Ip0}.

As anticipated based on the precision of the rough fit to $xg$, our
analytic estimate \nr{analytic} reproduces the ``exact'' numerical
result (with gluons only, no shadowing) to about 10\% accuracy near
$p_0=2$ GeV at $\roots=5500$ GeV. An improved accuracy would require a
better fit to $xg(x,Q^2)$ in the regions of large $y_2$(large $x_1$).
The terms ${\cal O}(\sqrt{x_{T0}/2})$, now neglected, contribute only at the
level of a few percent at $\roots=5500$ GeV, $p_0=2$ GeV, and 
at a level of 10\% at $\roots=200$ GeV, $p_0=1$ GeV.
Since the main emphasis here is to understand the origin of 
the scaling exponents, we leave the overall normalization as a 
rough estimate.

To extract the $A$ and $\roots$ scaling exponents for the saturation
scale analytically, we introduce a second parameter $\xi$ by noting
that the complicated $p_0$ dependence of the product $\alpha_s^2f$
in \nr{analytic} can, in the relevant range, 
to good accuracy be represented by a power:
\be
\alpha_s(p_0^2)\sqrt{f(\log(p_0^2/\lQCD^2))}
\approx D(\lQCD/p_0)^\xi,
\la{ksai}
\ee
where $D=0.775$ and $\xi=0.444$ (with $\lQCD=0.2$ GeV and $N_f=4$).  The
accuracy of this approximation is within 1.5 \% in the region
$p_0=1\dots 2$ GeV.

Also the first $p_T$-moment of the $p_T$-distribution can be computed
by using the same sequence of approximations as above. The result is 
\ba
\sigma_{\rm pQCD}\langle E_T\rangle &\approx&
{d\sigma\over dy} \langle p_T\rangle\bigg|_{y=0\atop p_0}
\approx \int_{p_0}^\infty2\pi p_Tdp_T {d\sigma\over dy d^2p_T}\cdot p_T\\
&\approx&
2K\frac{9\pi}{2}B(\delta,\frac{1}{2}) a(\delta)\cdot
2f(\log\frac{p_0}{\lQCD})\frac{\alpha_s^2(p_0)}{p_0}[xg(x_{T0},p_0^2)]^2,
\ea
where now the same function $f$ appears as in Eq. \nr{analytic} but 
with a different argument. For the average $p_T$, we thus get
\be
\langle p_T\rangle\bigg|_{y=0\atop p_0} \approx
\frac{2f(\log (p_0/\lQCD))}{f(\log (p_0^2/\lQCD^2))}\cdot p_0
\approx F \cdot\bigg(\frac{p_0}{\lQCD}\bigg)^{\eta} \cdot p_0
\la{ETanalytic}
\ee
where in the last step the power law approximation again holds in the region 
$p_0=1\dots 2$ GeV and $\eta=0.0624$ and $F=1.399$.

\section{The scaling exponents}

We can now apply the analytic approximations \nr{analytic} and
\nr{ksai} to the minijet cross section \nr{minijetsigma} in the final
state saturation condition \cite{ekrt} for central $A$+$A$ collisions:
\be N(p_0)=T_{AA}(\bfb=0)\,2\sigma_\rmi{pQCD}(p_0)=p_0^2R_A^2.
\la{psatcond} \ee Saturation is a dynamic phenomenon and, in the weak
coupling limit, there would be powers of $\alpha_s$ together with
various numerical constants in \nr{psatcond}. Taking a constant value
$\approx0.3$ for $\alpha_s$ the net effect in \nr{psatcond} is an
overall constant of about 1.  Even at the LHC one is most likely far
from the weak coupling region and we shall not keep the coupling
constant dependence in the right hand side of \nr{psatcond}
explicitly. 
This approximation is in agreement with RHIC data.  Note,
however, that $\alpha_s$ is kept in \nr{analytic}.

Using $T_{AA}(0)=A^2/(\pi R_A^2)$ with $R_A\approx 1.12 A^{1/3}$ and
Eq. \nr{analytic} in the power-law approximation \nr{ksai}, 
one finds that the solution of \nr{psatcond} is 
\be 
\psat\approx
\bigg[\frac{3\sqrt{B(\delta,\frac{1}{2})a(\delta)}}{1.12^2\cdot ({\rm
fm}\cdot {\rm GeV})^2} \cdot C_0(\frac{1}{2})^{\delta}\cdot
D\cdot\lQCD^{\xi} \bigg]^{1/(2+\xi)} K^{1/(4+2\xi)}
A^{1/(6+3\xi)}{\roots}^{\delta/(2+\xi)}, 
\la{psat} 
\ee 
where now the origin of each factor can be easily traced down. 
The exponent $\delta$ comes from the behaviour of the gluon structure 
function in Eq. \nr{simplefit}
whereas the exponent $\xi$ originates from the running
of the strong coupling constant in Eq. \nr{ksai}.
The numerical value of the constant in front of the
$K$-factor is 0.1625. It is also understood that $\psat$, $\lQCD$
and $\roots$ are in units of GeV.  We have also kept $K$ separate to
show how $\psat$ and, especially, $N(\psat)$ depend on it. The initial
multiplicity of produced gluons at saturation, 
$N(\psat)=\psat^2R_A^2$, then is
\be 
N(\psat) \approx 0.850\cdot K^{1/(2+\xi)}
A^{(6+2\xi)/(6+3\xi)}{\roots}^{2\delta/(2+\xi)}
\la{npsat}
\ee

Note that the dependence on the $K$ factor is rather weak, 
$N(\psat)\sim K^{0.41}$ --  instead of $\sim K$.
Substituting the numerical values for the coefficients $C_0$ and $D$,
and for the exponents $\delta$ and $\xi$ as discussed above, and $K=2$
as in \cite{ekrt}, we obtain the following scaling laws:
\ba
\psat 	&\approx& 0.187 A^{0.136}\roots^{\,0.205},
\la{analyticpsat}
\\
N(\psat)&\approx& 1.13 A^{0.939}\roots^{\,0.409}.
\la{analyticNsat}
\ea

Since the numerical results in Eqs. \nr{psatfit} and \nr{npsatfit}
contain shadowing, which is
not included in the analytic estimates above, 
we should compare the scaling laws obtained above
with the ones obtained numerically without shadowing (all parton flavours 
included):  
\ba
\psat&=&0.193\,A^{0.137}\,\roots^{\,0.204},
\la{psatfitnoshad}\\
N(\psat)&=&1.20\,A^{0.941}\roots^{\,0.408}.
\la{npsatfitnoshad}
\ea
The agreement is good and 
we have thus analytically understood how these scaling laws arise.

The numerical result for the CTEQ5 set \cite{cteq5} (no shadowing) is
$\psat=0.208\,A^{0.141}\,\roots^{\,0.192}$,
$N(\psat)=1.37\,A^{0.952}\roots^{\,0.384}$. The somewhat slower
$\roots$ dependence follows from a somewhat slower evolution in
this set, $\delta\approx0.47$.

Based on Eqs. \nr{psatcond} and \nr{analytic}
the multiplicity of produced gluons at saturation can 
also be cast in the form 
\be
N(\psat)\approx \sqrt{K9B(\delta,{\scriptstyle\frac{1}{2}})a(\delta)}
\sqrt{f(\log{(\psat^2/\lQCD^2)})}\,
\alpha_s(\psat^2)\,A\,xg(x_{\rm Tsat},\psat^2)
\la{Nsatclosed}
\ee
Thus we see that the initial multiplicity of produced 
gluons directly probes the gluon distribution at the 
saturation scale, as derived in \cite{bm} for initial state saturation.
The powers of $\alpha_s$ are not the same because they differ already
in the saturation condition \nr{psatcond}.

Nuclear shadowing effects can also be discussed in the analytic
approximation. Overall 
they are a fairly small correction to the results above: 
the numerical evaluation of $N(\psat)$  with the EKS98 shadowing \cite{EKS98}
shows a 16 \% reduction at $\roots=5500$ GeV and a 7 \% reduction
at $\roots=200$ GeV for $A=208$. For smaller nuclei the effects are 
smaller. Shadowing obviously slightly decreases the effective 
exponent $\delta$ in an $A$-dependent way. The dependence of the factor 
$\sqrt{B(\delta,{\scriptstyle\frac{1}{2}})a(\delta)}$ on $A$ remains, 
however, small. Disregarding the few percent effects from the factor 
$\sqrt{f(\log{(\psat^2/\lQCD^2)})}$, 
we arrive at the following simple scaling for the multiplicity of produced 
gluons at saturation
\be 
N(\psat)\sim A \alpha_s(\psat^2)xg_A(x_{\rm Tsat},\psat^2),
\la{nsatclosedshad}
\ee 
where now $\psat$ is from Eq. \nr{psatfit} and $xg_A$ is the shadowed 
gluon distribution per nucleon. This result is tested
against a full calculation of Eq. \nr{npsatfit} in Fig. 3.
The agreement with the numerically obtained results is good in the 
scalings with both $\roots$ and $A$, especially at large $\roots$ and 
large $A$.\footnote{The slight 
kink in the curves with shadowing originates from taking the shadowing 
to be scale independent at $\psat$ smaller than the minimum $Q$ in 
the EKS98 parametrization.} 
If the initial state multiplicity is directly proportional to the
final state multiplicity, the measured charged particle multiplicity
then directly probes the nuclear gluon distributions at the (final
state) saturation scale.

\begin{figure}[hbt]
\vspace{-1cm}
\hspace{0cm}
\centerline{\hspace{0cm}\epsfysize=9.5cm\epsffile{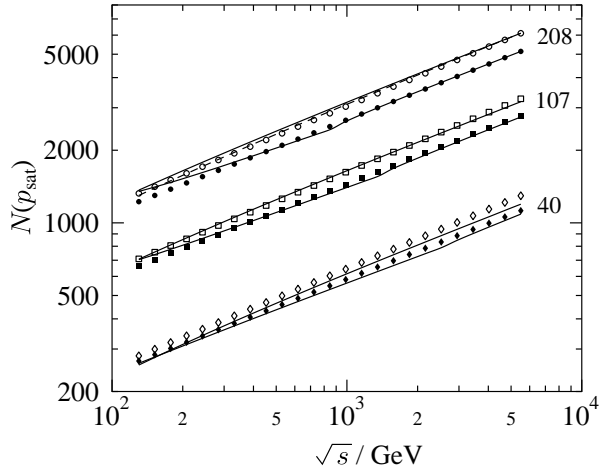}}
\vspace{-2cm}
\caption[a]{\small The initial multiplicity of produced gluons
$N(\psat)$ at saturation as a function of $\roots$.  The symbols are
the numerically obtained fits, Eq. \nr{npsatfit} with shadowing
included (filled symbols), and Eq. \nr{npsatfitnoshad} with no shadowing
(open symbols).  Circles, boxes and diamonds stand for $A=208$, 107
and 40, correspondingly. The solid curves are the prediction of
Eq. \nr{nsatclosedshad}, with $\psat$ computed from the numerical fits,
Eqs. \nr{psatfit} and \nr{psatfitnoshad} and with GRV94LO gluon densities 
with and without shadowing. 
The solid curves are normalised to the numerically obtained result
for $A=208$ at $\roots=5500$ GeV. The dashed curve shows the small effect of
the term $f(\log(\psat^2/\lQCD^2))$.
 }  \la{nsatfig}
\end{figure}

Using the power-law approximation of Eq. \nr{ETanalytic}, we obtain
the average initial transverse energy per produced particle
\ba
\frac{E_T(\psat)}{N(\psat)} &=& \frac{T_{AA}(0)\sigma_{\rm pQCD}
\langle E_T\rangle}{T_{AA}(0)2\sigma_{\rm pQCD}} \approx \langle p_T\rangle\bigg|_{y=0, \psat} \approx 1.546\bigg(\frac{\psat}{\gev}\bigg)^{\eta}\cdot \psat.
\ea
From here, using the analytic approximation for $\psat$ from Eq. \nr{psat}
and $N(\psat)=\psat^2 R_A^2$, we get
\ba
E_T(\psat)&=& 0.191  K^{(3+\eta)/(4+2\xi)} A^{2/3 + (3+\eta)/(6+3\xi)} \roots^{\delta(3+\eta)/(2+\xi)}\\
&=&0.191 K^{0.627} A^{1.08}\roots^{0.626}.
\ea

At saturation, the initial number and energy densities become
\be
n_i={N(\psat)\over V(\psat)}={1\over\pi}\psat^3,\quad
\epsilon_i={E_T(\psat)\over V(\psat)}=\psat^4{1.546\over\pi}\bigg(\frac{\psat}{\gev}\bigg)^{0.0624}
,\,\,\, 
V(\psat) = \pi R_A^2/\psat.
\ee

For a thermalised system of massless bosons at an energy density
$\epsilon_{\rm th}=\epsilon_i$, the ratio initial energy per particle
can be written as \be \frac{\epsilon_{\rm th}}{n_{\rm th}}= 2.7 T =
2.7 \bigg(\frac{30\epsilon_i}{16\pi^2}\bigg)^{1/4} \approx
\frac{E_T(\psat)}{N(\psat)} \cdot 0.97
\bigg(\frac{\psat}{\gev}\bigg)^{-0.047}, \ee which is indeed very
close to the computed ratio $E_T(\psat)/N(\psat)=\epsilon_i/n_i$,
independent of $A$ and $\roots$ \cite{ekrt}. Since the system looks
thermal from the point of view of the average quantities, rapid
thermalisation is plausible. This is to be contrasted with the
classical field approach \cite{McLV} where it is found \cite{KV1,KV2}
that the ratio $E_T/N$ is approximately three times larger, and
therefore one might expect that thermalisation takes longer to
achieve. On the other hand, from the analytic classical field
calculation of Kovchegov \cite{K1} one infers that the $\epsilon/n$
ratio is very close to the thermal one \cite{K2} and, again, rapid
thermalisation would be expected.

\section{Final vs. initial state saturation}

Usually saturation is discussed as a small-$x$ property of parton
distribution functions. The above computations have been formulated
referring to saturation of final state partons. One clearly 
has to understand the relation between these two approaches.

An initial state saturation scale $\qsat$ can naturally be defined
\cite{gm} as the gluon transverse area density including all gluons
with $x>x_\rmi{sat}=2\qsat/\roots$: 
\be {N_g\over\pi R_A^2}={A\over\pi
R_A^2}\int_{x_\rmi{sat}}^1\,dy\, g(y,\qsat^2)={A\over\pi
R_A^2}C_0\delta^{-1}(\qsat /{\rm{GeV}})^\delta x_\rmi{sat}^{-\delta}
={1\over\pi}\qsat^2, \la{psatinit} \ee 
where the second equality was
obtained by approximating $xg(x,Q^2)=C_0(Q/\gev/x)^\delta$ as before in
Eq.\nr{simplefit}. This equation is geometric and thus analogous to
the saturation condition \nr{psatcond}. In the parametric weak
coupling limit also this equation would contain various group theory
factors and powers of the coupling constant. As already discussed, we
set them equal to unity in this work. As noted below
Eq. \nr{Nsatclosed}, the powers of $\alpha_s$ in the saturation
condition will affect the parametric dependence of e.g. multiplicity on
$\alpha_s$.

Approximating the gluon distribution as earlier and solving $\qsat$
from Eq. \nr{psatinit} gives almost the same A- and $\roots$-scaling
exponents as in \nr{psat} for $\psat$, only the constant is somewhat
different and the $K$-factor is absent: 
\ba
\qsat&=&\frac{1}{{\rm{fm\cdot GeV}}}(\frac{C_0}{2^{\delta}\delta})^{1/2}
(\frac{A}{R_A^2})^{1/2}
\roots^{\delta/2}\sim A^{1/3}\roots^{\delta/2},
\ea
where $\qsat$ and $\roots$ are in units of GeV and $R_A$ in fm.
Note that it is essential that $\qsat$ be both in the lower
limit and on the right hand side of Eq.\nr{psatinit}. The
constant anyway is not uniquely defined, since the
lower limit in \nr{psatinit} is not unique. Thus
\be
\psat\approx\qsat.
\ee
In fig. \nr{initsat} we plot the determination of the saturation scale
using both final multiplicity and initial gluon distribution. From
this figure we can see the small difference between the
$\roots$-scaling in $\psat$ and $\qsat$. 

In the initial state saturation -picture the multiplicity $N$ of produced
gluons is expected to be proportional to $N_g$ in Eq. \nr{psatinit}, i.e.
\be
N\sim N_g\sim\qsat^2 R_A^2 
\ee
This relation is described in terms of the ``parton liberation"
constant in \cite{mueller_c}, and has been confirmed in the lattice
simulations of the classical fields \cite{KV2}.

These results suggest that finding dynamical saturation of gluon
distribution functions in a nucleus, one should also find saturation
of produced gluons, the two phenomena are intimately related.

\begin{figure}[hbt]
\hspace{0cm}
\centerline{\hspace{1cm}\epsfysize=9.5cm\epsffile{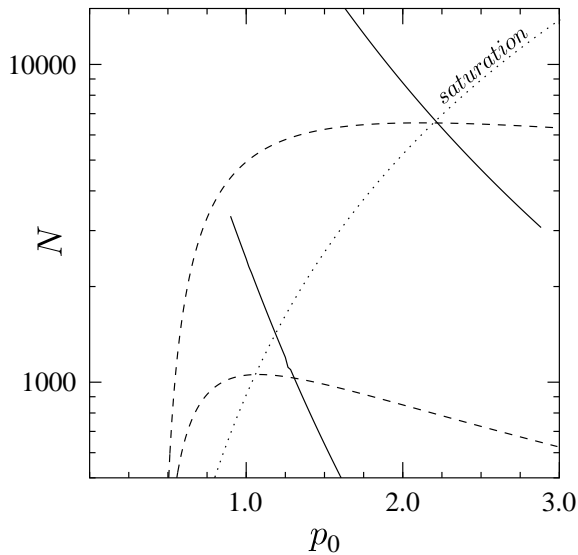}} 
\vspace{-2.5cm}
\caption{\small Solution of the saturation scale
obtained by using the final multiplicity (solid lines) and
Eq. \nr{psatcond}, or the initial gluon multiplicity (dashed lines)
and Eq. \nr{psatinit}. The saturation scale is given by the
intersection of these curves with the dotted line 'saturation'
corresponding to $p_0^2R_A^2$. Upper two curves correspond to LHC energy while
the lower two correspond to the full RHIC energy. Neither multiplicity
contains shadowing and $A=197$ on all curves. The dashed curves
could be compared with those in Fig.2 of \cite{gm}.
}
\la{initsat}
\end{figure}

\section{Local saturation}

In \cite{ekt} the criterion \nr{psatcond} was generalized to a local
condition for transverse saturation of produced gluons in a collision
with impact parameter $\bfb$: 
\be
\frac{dN}{d\bfs}=T_A(\bfs)T_A(\bfb-\bfs)2\sigma_\rmi{pQCD}(\psat)=
\frac{1}{\pi} \psat^2(\bfb,\bfs), 
\la{localpsatcond} 
\ee where $\bfs$
is the transverse coordinate; see also \cite{pirner}.
Using Eq.\nr{analytic} in
\nr{localpsatcond} one finds that exactly the same $A$ and $\sqrt s$
scaling exponents are obtained as from Eq.\nr{psatcond}, and that the
dependence on impact parameter and transverse coordinates is isolated
into a product of nuclear density functions with a $\xi$-dependent
exponent: 
\be \psat^2(\bfb,\bfs)\sim K^{1/(2+\xi)}\sqrt
s^{2\delta/(2+\xi)} [T_A(\bfs)T_A(\bfb-\bfs)]^{1/(2+\xi)}
\la{localpsat} 
\ee 
and 
\be N(\bfb)\sim \frac{1}{\pi}K^{1/(2+\xi)}
\sqrt s^{2\delta/(2+\xi)} \int
d^2\bfs[T_A(\bfs)T_A(\bfb-\bfs)]^{1/(2+\xi)}.  
\la{localN} 
\ee 
Again,
$\delta=0.5$ and the parameter $\xi$ as given by Eq.\nr{ksai} reproduce 
the behaviour of $N(\bfb)$ obtained in the
numerical computation in \cite{ekt}. 

With our ansatz \nr{simplefit} for $xg(x,Q^2)$, 
and neglecting the $p_0$-dependence of $f(\log p_0^2/\lQCD^2)$
in Eq. \nr{analytic},
Eq. \nr{localpsatcond} can also be cast into the form
\be
p_\rmi{sat}^2(\bfb,\bfs)\sim K^{1/2}[T_A(\bfs)T_A(\bfb-\bfs)]^{1/2}
\alpha_s xg(x,\psat^2(\bfb,\bfs)) 
\la{psatwithxg} 
\ee
and, consequently, 
\be
N(\bfb)\sim K^{1/2}\int d^2\bfs[T_A(\bfs)T_A(\bfb-\bfs)]^{1/2}
\alpha_s xg(x,\psat^2(\bfb,\bfs)),
\la{Nwithxg} 
\ee
where $x=2\psat(\bfb,\bfs)/\roots$.

Eqs.\nr{psatwithxg} and \nr{Nwithxg} permit us to comment on the
relation to \cite{kharzeevnardi}, where it was 
postulated that the average (over $\bfs$) saturation scale 
$Q_\rmi{s}^2$ (at fixed $b$) be proportional to the average 
(over $\bfs$) transverse density of participating nucleons, 
\be
Q_\rmi{s}^2=\frac{8\pi^2\alpha_sN_\rmi{c}}
{N^2_\rmi{c}-1}xg(x,Q_\rmi{s}^2)
\frac{\rho_\rmi{part}}{2}. 
\ee
This leads to a total multiplicity, at fixed $b$,
\be 
\frac{dN}{d\eta}=cN_\rmi{part}xg(x,Q_\rmi{s}^2), 
\ee
with $c\gsim1,\,x=2Q_\rmi{s}/\sqrt s$. 
Also, the quantity $N/N_\rmi{part}$ is a
slightly increasing function of $N\rmi{part}$ due to 
assumed scale evolution of the gluon structure function 
of the type $xg(x,Q^2_\rmi{s})\sim 
\ln(Q_\rmi{s}/\lQCD)$. The difference 
between \cite{kharzeevnardi} and \cite{ekt} can be traced
down to two points:
First, to a slightly different dependence of the saturation scale on the
transverse coordinate originating from $\rho_\rmi{part}
\leftrightarrow [T_A(\bfs)T_A(\bfb-\bfs)]^{1/2}$. Second, to a different
order of averaging to obtain $N(\bfb)$, namely $N_\rmi{part}xg(
x,Q_\rmi{s}^2)\leftrightarrow \int
d^2\bfs[T_A(\bfs)T_A(\bfb-\bfs)]^{1/2}\alpha_s xg(x,\psat^2(\bfb,\bfs))$.

\section{Discussion}
We have here shown how the $A$- and $\roots$-scaling
exponents and the overall magnitude
of various global quantities in ultrarelativistic
$A+A$ collisions, numerically computed in \cite{ekrt}, can be
simply related to two parameters, $\delta$ and $\xi$. The
former (Eq.\nr{simplefit}) is related to
the $x^{-\delta}$ behaviour at small-$x$ 
of the gluon distribution function at an effectively $x$-dependent
saturation scale. Due to the interdependence of $x$ and $Q$
this is not the standard BFKL exponent
describing small-$x$ behaviour at fixed scale $Q$. The parameter $\xi$
(Eq.\nr{ksai}) approximates a complicated function containing
$\alpha_s$ by a power.
All the exponents are
accurately reproduced by $\delta\approx0.5,\,\,\xi=0.44$. 
The consequences of initial
and final state saturation were also shown to be quantitatively
similar.

One may note the following:
\bi
\item The $A$-dependence of $N$ is not that of independent hard
scatterings ($\sim A^{4/3}$), nor that of the saturation model
with scaling cross section $\sigma_\rmi{pQCD}\sim 1/p_0^2$
($\sim A$) but, 
due to powerlike non-scaling of $\sigma_\rmi{pQCD}$
even slower, $\sim A^{(6+2\xi)/(6+3\xi)}\sim A^{0.94}$. 
This is so even without shadowing, which further 
slightly reduces the
exponent. A qualitative effect of this is that in the study
of multiplicity per 0.5 times number of participants at some
impact parameter $\bfb$ (which is the number to use to compare
A+A data at various $\bfb$ with pp collisions) one obtains
a curve decreasing very slowly with $N_\rmi{part}$ \cite{wg}. 
In fact, using the
simple estimate $A_\rmi{eff}=0.5N_\rmi{part}(\bfb)$, Eq.\nr{npsat}
implies that
\be
{dN_\rmi{ch}/dy\over0.5N_\rmi{part}}=\fr23\cdot 0.9 \cdot
0.8503\cdot K^{1/(2+\xi)}
(0.5N_\rmi{part})^{-\xi/(6+3\xi)}{\roots}^{2\delta/(2+\xi)}.
\la{nchnpart}
\ee
The decrease is thus very slow, $\sim N_\rmi{part}^{-0.06}$
for $\xi=0.444$. A more accurate analysis, using a local
saturation condition \cite{ekt}, leads to a virtually constant
($\bfb$-independent) ratio at RHIC and slightly increasing ratio at LHC, 
as shown in Fig.\ref{nch}.

\begin{figure}[!htb]
\hspace{0.5cm}
\epsfysize=12cm
\centerline{\epsffile{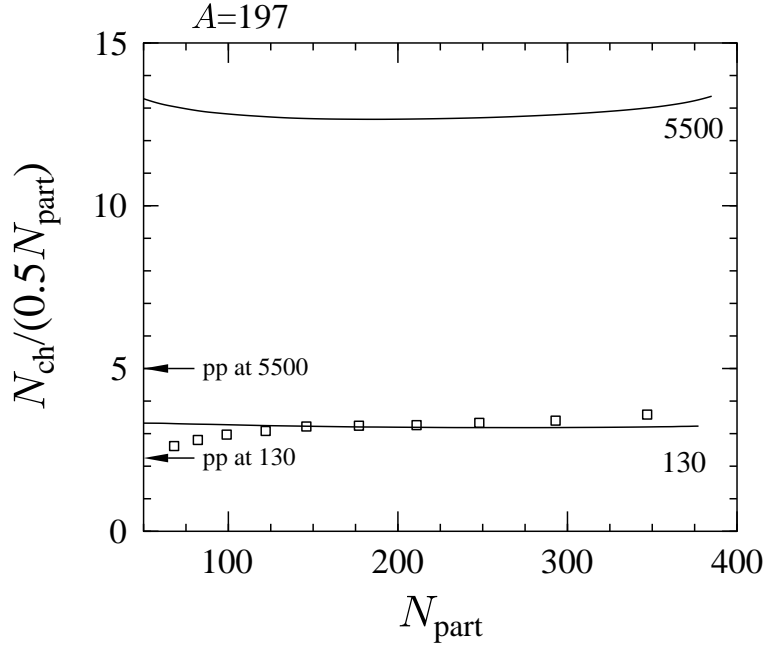}}
\vspace{-2cm}
\caption[a]{\protect \small Rapidity density of charged particles
near $y=0$ per 0.5 times the number of 
participants at LHC and RHIC energies computed using the local
saturation criterion in \cite{ekt}. RHIC data at $\roots$ = 130 GeV
\cite{phenix} (open squares) and p+p rapidity densities 
(then $N_\rmi{part}=2$; arrows) are also shown. 
Large (small) $N_\rmi{part}$
corresponds to central $\bfb=0$ (peripheral, $\bfb\to2R_A$)
collisions.
}
\la{nch}
\end{figure}

\item The energy dependence of $N$ and also of the ratio 
$N_\rmi{ch}/(0.5N_\rmi{part})$ is the powerlike 
$\roots^{\,2\delta/(2+\xi)}\approx\roots^{\,0.41}$ for 
$\delta=0.5,\, \xi=0.44$. This simple power behaviour follows
from the numerically accurate power approximation \nr{ksai}.
This $\roots$ dependence is roughly verified at 
RHIC for $\roots=56,130$
GeV and a new check is soon obtained with data at $\roots=200$
GeV. At RHIC energies A+A collisions ($A=197$), have
$N_\rmi{ch}/(0.5N_\rmi{part})\approx3$, clearly but not
strikingly larger than the value of $\approx$ 2 for p+p collisions.
At LHC the increase would be from 5 for p+p to 
about 13 for A+A (Fig.\ref{nch}),
a really striking effect, which will directly probe the behaviour of the 
nuclear gluon densities at small values of $x$.

\item As noted 
previously, the powers of $\alpha_s$ appearing in the
formulae for the multiplicity of produced gluons, Eqs. \nr{Nsatclosed}
and \nr{nsatclosedshad}, 
will be affected by additional powers of $\alpha_s$ in the saturation
condition \nr{psatcond}, which will appear in the weak coupling limit but
which were replaced by constants in this study covering a
limited energy range.
However, it is interesting to note that the
exponent $\delta$ of the structure function appears only (at least
when shadowing is neglected) in the $\roots$-scaling. The $A$-scaling
of the multiplicity depends only on the exponent $\xi$, which is
related to $\alpha_s$. Including explicitly additional powers of
$\alpha_s$ in the saturation condition \nr{psatcond}, the $\xi$
dependence of $A$-scaling would change, and therefore the experimental
measurement of $A$-scaling of the multiplicity would be a measurement
of the actual form of the saturation criterion itself. 
Inclusion of, say, a factor $1/\alpha_s^2$, would make the
$\alpha_s$ dependence of multiplicity of Eq. \nr{Nsatclosed} and that
of \cite{bm} consistent with each other. It will also be interesting
to study the relation to the self-screened parton cascades \cite{emw}.
We emphasize again, however, that the purpose of this paper was to 
understand the scaling laws
obtained numerically in \cite{ekrt}, where no explicit powers of
$\alpha_s$ were considered in the saturation condition.

\ei

{\bf Acknowledgements} We thank M. Gyulassy, D. Kharzeev, 
Yu. Kovchegov and X.-N. Wang  for discussions. 
Financial support from the Academy of Finland 
(grants No. 43989 and 773101) is gratefully acknowledged.

\end{document}